\documentclass[onecolumn]{revtex4}

\usepackage{graphicx}
\usepackage{dcolumn}
\usepackage{bm}
\usepackage{color}
\usepackage{amssymb,amsmath}

\input epsf

\begin{document}

\title{\Large{Observational Constraints of Homogeneous Higher Dimensional Cosmology with Modified Chaplygin Gas}}

\author{\bf Chayan Ranjit$^1$\footnote{chayanranjit@gmail.com}, Shuvendu
Chakraborty$^2$\footnote{shuvendu.chakraborty@gmail.com} and Ujjal
Debnath$^3$\footnote{ujjaldebnath@yahoo.com}}
\affiliation{$^{1,2}$Department of Mathematics, Seacom Engineering College, Howrah - 711 302, India.\\
$^3${Department of Mathematics, Bengal Engineering and Science
University, Shibpur, Howrah-711 103, India.} }

\date{\today}

\begin{abstract}
In this work, we have considered the flat FRW model of the
universe in $(n+2)$-dimensions filled with the dark matter
(perfect fluid with negligible pressure) and the modified
Chaplygin gas (MCG) type dark energy. We present the Hubble
parameter in terms of the observable parameters $\Omega_{m0}$,
$\Omega_{x0}$ and $H_{0}$ with the redshift $z$ and the other
parameters like $A$, $B$, $C$, $n$ and $\alpha$. From Stern data
set (12 points), we have obtained the bounds of the arbitrary
parameters by minimizing the $\chi^{2}$ test. The best-fit values
of the parameters are obtained by 66\%, 90\% and 99\% confidence
levels. Now to find the bounds of the parameters and to draw the
statistical confidence contour, we first fixed three parameters
$C, n, \alpha$ and then fixed the three parameters $A, n, \alpha$.
In the first case we find the bounds of $(A, B)$ and draw the
contour between them for 4D$(n=2)$, 5D$(n=3)$ and 6D$(n=4)$. In
the second case we fixed three different values of A as 1, $1/3$,
$-1/3$ to find the bounds of $(B, C)$ and draw the contour between
them. Here the parameter $n$ determines the higher dimensions and
we perform comparative study between three cases : 4D $(n=2)$, 5D
$(n=3)$ and 6D $(n=4)$ respectively. Next due to joint analysis
with BAO observation, we have also obtained the bounds of the
parameters ($A,B$) by fixing some other parameters $\alpha$ and
$A$ for 4D, 5D and 6D.
\end{abstract}

\maketitle

\section{\normalsize\bf{Introduction}}

In recent cosmological research work, the theoretical models and
range of the cosmological parameters are tested continuously by
the combination of different observational astrophysical data.
Before achieving Supernova data, it seems that the universe may be
occupied by energy density which is very well distributed over
large scales \cite{Maddox}. In 1992, Cosmic Background Explorer
(COBE) \cite{paddy3,Bond1} data suggested that the spectrum of
Standard Cold Dark Matter (SCDM) should be modified and proved the
necessity of existence of the cosmological constant. The existence
of non-zero cosmological constant $\Lambda$ (which has the
equation of state $w_{\Lambda}=-1$) was supported in 1996
\cite{Paddy4}.
 The different cosmological observation of SNeIa
\cite{Perlmutter,Perlmutter1,Riess,Riess1}, large scale redshift
surveys \cite{Bachall,Tedmark}, the measurements of the cosmic
microwave background (CMB) \cite{Miller,Bennet} and WMAP
\cite{Briddle,Spergel} anticipate that our present universe which
is expanding with acceleration, preceded by a period of
deceleration. The mysterious observational facts were not
explained by the standard big bang Cosmology with perfect fluid.
Thus to integrate the recent prediction from observational
cosmology, a modification is necessary in the matter sector of the
Einstein Gravity. The unknown candidate which is responsible for
this accelerating scenario, has the property that the positive
energy density and sufficient negative pressure, known as dark
energy \cite{Paddy,Sahni}. The scalar field or quintessence
\cite{Peebles} produce sufficient negative pressure to provoke
acceleration in which the kinetic term is dominated by the
potential. From recent cosmological observations including
supernova data \cite{Riess2} and measurements of cosmic microwave
background radiation (CMBR) \cite{Spergel1} it is evident that in
$\Lambda$CDM model the Universe is made up of about $\sim$ 26\%
matter (baryonic + dark matter) and $\sim$ 74\% of a smooth vacuum
energy component.\\

For $z>0.01$, the TONRY data set with the 230 data points
\cite{Tonry} together with the 23 points from Barris et al
\cite{Barris} are valid. Another data set named the ``gold''
sample (see \cite{Riess1}) contains 156 points, which includes the
latest points observed by HST and this covers the redshift range
$1 < z < 1.6$. Recently the CMBR data (for recent WMAP results,
see \cite{Spergel}) strongly support
$\Omega_{\Lambda}+\Omega_{m}=1$ for FRW universe in Einstein
gravity. Also in $2007$, Choudhury et al \cite{Paddy1} showed that
the best-fit value of $\Omega_{m}$ for flat model was $0.31\pm
0.08$. \\

One of the most effective candidate of dark energy having positive
energy density and negative pressure is Chaplygin gas whose EOS is
given by $p=-B/\rho$ \cite{Kamenshchik} with $B>0$. Later, it has
been generalized to the form $p=-B/\rho^{\alpha}$
\cite{Gorini,Bento} and thereafter modified to the form
$p=A\rho-B/\rho^{\alpha}$ \cite{Debnath}, which is known as
Modified Chaplygin Gas (MCG). The Modified form of Chaplygin Gas
go with the 3 year WMAP and the SDSS data with the choice of
parameters $A =0.085$ and $\alpha = 1.724$ \cite{Lu} which are
improved constraints than the previous
ones ($-0.35 < A < 0.025$) \cite{Jun}.\\

The drawback of the gravitational force has been successfully
explained by proposing the existence of more than three special
dimensions \cite{Poppenhaeger}. Today, the existence of extra
dimensions \cite{Arkani, Panigrahi1} are supported by large number
of promising model and theories. The solutions for MCG in
$(n+2)$-dimensional FRW Cosmology are given in section II. Here,
the table of $H(z)$ and $\sigma(z)$ is presented for different
values of $z$. The $\chi^{2}$ minimum test for best fit values of
parameters are investigated with Stern and Stern+BAO joint data
analysis in section III. Finally, some observational conclusions are drawn.\\

\section{\bf{Basic Equations and Solutions for MCG in Higher Dimensional FRW Cosmology}}

We consider the $(n+2)$ dimensional flat $(k=0)$, homogeneous and
isotropic universe described by FRW metric is given by
\cite{Chatterjee, Utpal}

\begin{equation}
ds^{2}=dt^{2}-a^{2}(t)[dr^{2}+r^{2}dx_{n}^{2}]
\end{equation}
where $a(t)$ is the scale factor and
\begin{equation}
dx_{n}^{2}=d\theta^{2}+sin^{2}\theta_{1}d\theta^{2}_{2}+......+
sin^{2}\theta_{1}sin^{2}\theta_{2}...sin^{2}\theta_{n-1}d\theta^{2}_{n}
\end{equation}
 The modified Einstein's field equations in higher dimension are given by

\begin{equation}
\frac{n(n+1)}{2}\left(\frac{\dot{a}}{a}\right)^{2}=
\rho_{m}+\rho_{x}
\end{equation}
and
\begin{equation}
n\frac{\ddot{a}}{a}+\frac{n(n-1)}{2}\left(\frac{\dot{a}}{a}\right)^{2}=-p_{x}
\end{equation}

where $\rho_{m}$ is the energy density of the dark matter
(pressureless fluid) and $\rho_{x}$, $p_{x}$ are the energy
density and pressure of dark energy (choosing $8\pi G=c=1$).\\

Now we consider the Universe is filled with Modified Chaplygin Gas
(MCG) whose equation of state (EOS) is given by \cite{Debnath}

\begin{equation}
p_{x} = A\rho_{x}- \frac{B}{\rho_{x}^{\alpha}}~,~A > 0,~B
>0,~0\leq \alpha \leq 1
\end{equation}

We also assume that the dark matter and dark energy are separately
conserved. So the energy conservation equations in homogeneous
higher dimensional cosmology are

\begin{equation}
\dot{\rho}_{m}+(n+1)H \rho_{m}=0
\end{equation}
and
\begin{equation}
\dot{\rho}_{x}+(n+1)H(\rho_{x}+p_{x})=0
\end{equation}

where $H$ is the Hubble parameter defined as
$H=\frac{\dot{a}}{a}$. From the first conservation equation (6) we
have the solution of $\rho_{m}$ as

\begin{equation}
\rho_{m}=\rho_{m0}(1+z)^{n+1}
\end{equation}

where $\rho_{m0}$ is the present value of the density of matter
and $z=\frac{1}{a}-1$ is the cosmological redshift. From the
second conservation equation (7) we have the solution of the
energy density $\rho_{x}$ as

\begin{equation}
\rho_{x}=\left[\frac{B}{A+1}+\frac{C(1+z)^{(n+1)(\alpha+1)(A+1)}}{(A+1)}\right]^{\frac{1}{\alpha+1}}
\end{equation}

where $C$ is the integration constant which can be interpreted as
one of the contribution of dark energy. Now the equation (9) can
be written as

\begin{equation}
\rho_{x}=\rho_{x0}\left[\frac{B}{(1+A)C+B}+\frac{(1+A)C}{(1+A)C+B}
(1+z)^{(n+1)(\alpha+1)(A+1)}\right]^{\frac{1}{\alpha+1}}
\end{equation}
where $\rho_{x0}$ is the present value of the dark energy
density.\\

\section{\bf{Observational Data Analysis Mechanism}}

We now investigate the expected bounds of the theoretical
parameters by $\chi^{2}$ statistical best fit test with the basis
of $H(z)$-$z$ (Stern) \cite{Stern} and Stern+BAO \cite{Wu, Paul,
Paul1, Paul2, Paul3, Chak} joint data analysis. We also determine
the statistical confidence contours between any two parameters of
of MCG in FRW higher dimensional cosmology. To investigate the
bounds of model parameters here we consider Stern ($H(z)$-$z$)
data set with 12 data of $H(z)$-$z$ (Stern) given by \cite{Stern}

\[
\begin{tabular}{|c|c|c|}
\hline
  ~~~~~~~~~~~~$z$ ~~~~~~~~~~& ~~~~~~~~~~$H(z)$ ~~~~~~~~~~~~~& ~~~~~~~~~~~$\sigma(z)$~~~~~~~~~~~~\\
  \hline
  0 & 73 & $\pm$ 8 \\
  0.1 & 69 & $\pm$ 12 \\
  0.17 & 83 & $\pm$ 8 \\
  0.27 & 77 & $\pm$ 14 \\
  0.4 & 95 & $\pm$ 17.4\\
  0.48& 90 & $\pm$ 60 \\
  0.88 & 97 & $\pm$ 40.4 \\
  0.9 & 117 & $\pm$ 23 \\
  1.3 & 168 & $\pm$ 17.4\\
  1.43 & 177 & $\pm$ 18.2 \\
  1.53 & 140 & $\pm$ 14\\
  1.75 & 202 & $\pm$ 40.4 \\ \hline
\end{tabular}
\]
~~~~~~~~~~~~~~~~~~~~~~~~~~~~~~~~~~~~~~~~~{\bf Table 1:} $H(z)$ and
$\sigma(z)$ for different values of $z$. \vspace{6mm}

From the solutions (8) and (10) we now express the Hubble
parameter $H$ in terms of redshift  parameter $z$ and two
dimensionless density parameters $\Omega_{m0}=\frac{2
\rho_{m0}}{n(n+1) H_{0}^{2}}$ and
$\Omega_{x0}=\frac{2\rho_{x0}}{n(n+1) H_{0}^{2}}$ as follows:

\begin{equation}
H(z)=H_{0}\left[\Omega_{x0}\left[\frac{B}{(1+A)C+B}+\frac{(1+A)C}{(1+A)C+B}
(1+z)^{(n+1)(\alpha+1)(A+1)}\right]^{\frac{1}{\alpha+1}}+\Omega_{m0}(1+z)^{n+1}\right]^{\frac{1}{2}}
\end{equation}

This equation can be written in the form $H(z)=H_{0}E(z)$, where
$E(z)$ known as normalized Hubble parameter contains five model
parameters $A, B, C, n, \alpha$ beside the redshift parameter $z$.
Now to find the bounds of of the parameters and to draw the
statistical confidence contour (66\%, 90\% and 99\% confidence
levels) we first fixed three parameters $C, n, \alpha$ and then
fixed the three parameters $A, n, \alpha$. In the first case we
find the bounds of $A, B$ and draw the contour between them. In
the second case we fixed three different values of A as 1, $1/3$,
$-1/3$ to find the bounds of $B, C$ and draw the contour between
them. Here the parameter $n$ determines the higher dimensions and
we perform comparative study between three cases : 4D $(n=2)$, 5D
$(n=3)$ and 6D $(n=4)$ respectively.\\

\begin{figure}
\includegraphics[scale=0.5]{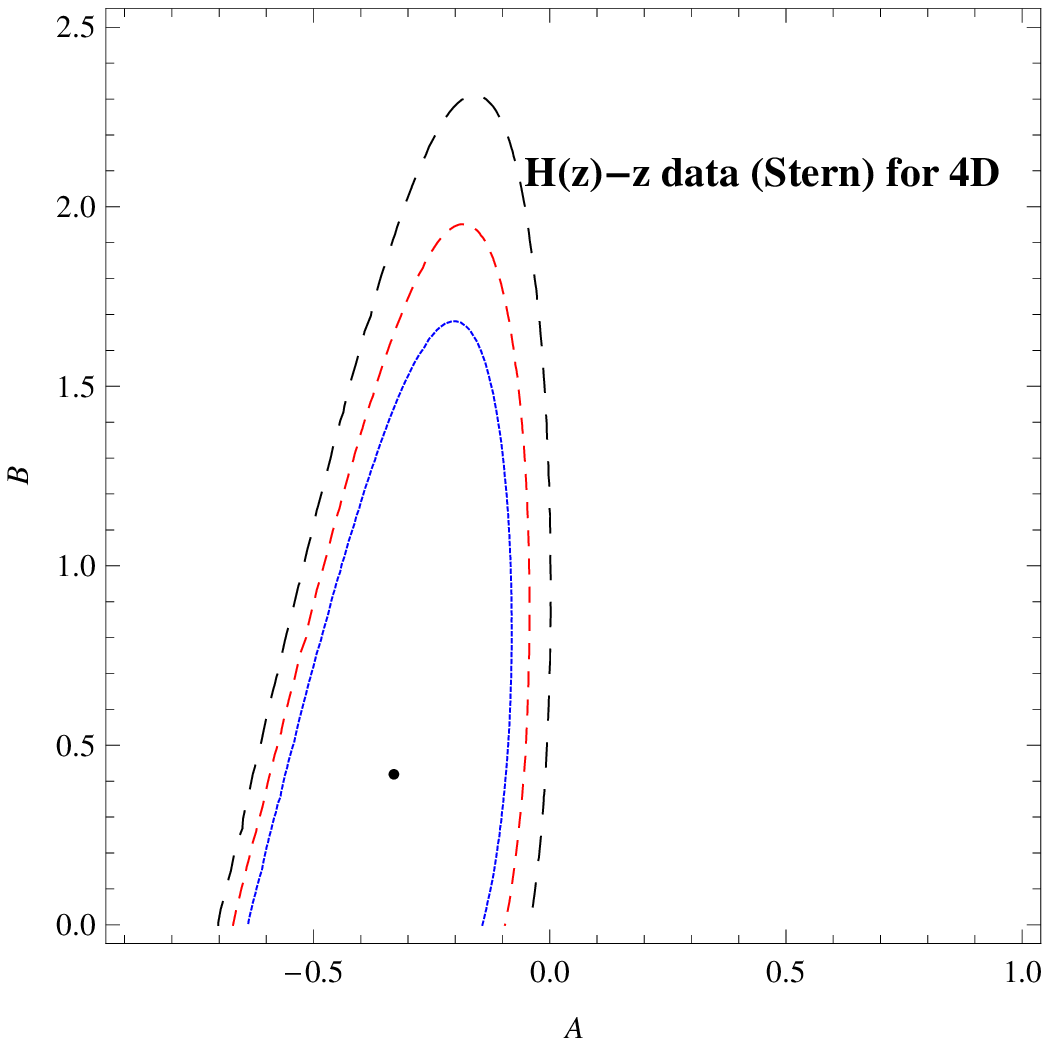}~~
\includegraphics[scale=0.5]{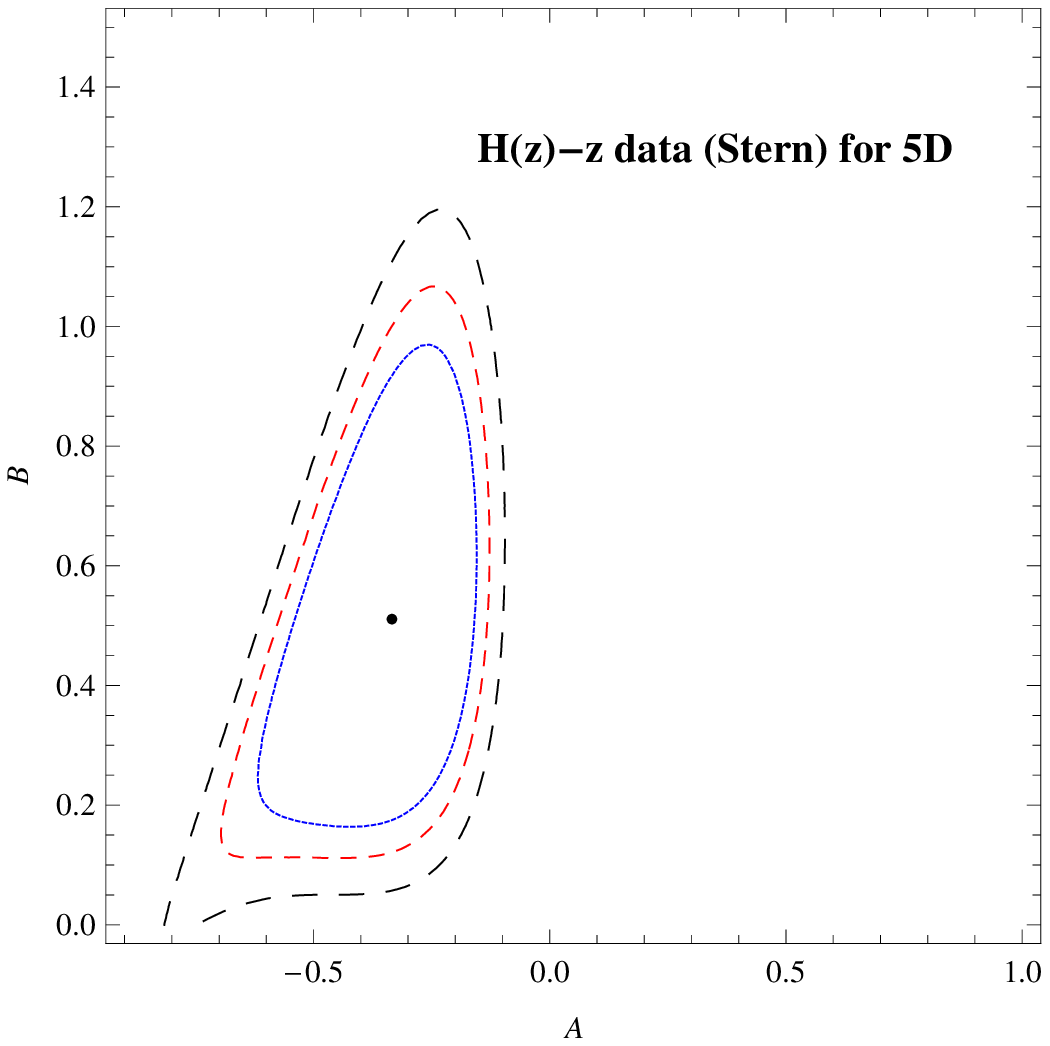}~~
\includegraphics[scale=0.5]{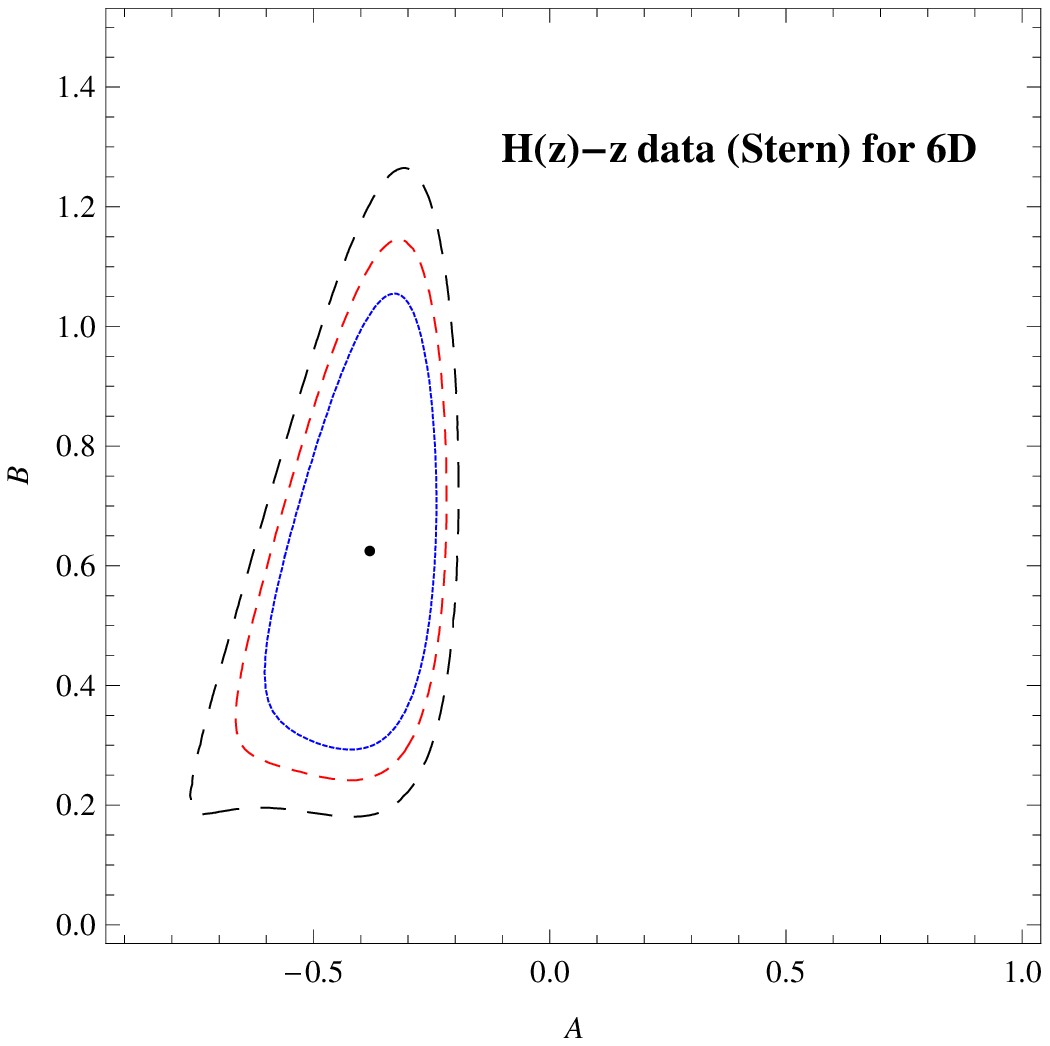}\\
\vspace{2mm}
~Fig.1~~~~~~~~~~~~~~~~~~~~~~~~~~~~~~~~~~~~~~~~~~~Fig.2~~~~~~~~~~~~~~~~~~~~~~~~~~~~~~~~~~~~~~~~~~~Fig.3\\

\vspace{4mm}

Figs. 1 - 3 show that the variation of $B$ against $A$ for
$\alpha= 0.0001$ and $C=1.5$ in 4D, 5D and 6D respectively for
different confidence levels. The 66\% (solid, blue), 90\% (dashed,
red) and 99\% (dashed, black) contours are plotted in these
figures for the $H(z)$-$z$ (Stern) analysis.

\vspace{4mm}
\end{figure}

\subsection{\bf{Stern ($H(z)$-$z$) Data Analysis}}

  Here we analyze the model parameters using twelve data \cite{Stern} of
 Hubble parameter for different redshift given by table 1. The
 corresponding error $\sigma (z)$ also available in the table.
 Here we perform $\chi^{2}$ test as a tool to find the minimum
 values of the parameters and draw the contours with three
 particular confidence limit. For this purpose we first define the
 $\chi^{2}$ statistic with 11 degree of freedom as
\begin{equation}
{\chi}_{Stern}^{2}=\sum\frac{(H_{E}(H_{0},A,B,C,n, \alpha
,z)-H_{obs})^{2}}{\sigma(z)^{2}}
\end{equation}

where $H_{E}$ and $H_{obs}$ are theoretical and observational
values of Hubble parameter at different redshifts respectively and
$\sigma(z)$ is the corresponding error as per Table 1. Since we
are interested to determine model parameters, $H_{0}$ is
considered as a nuisance parameter and can be safely marginalized.
We consider the observed parameters $\Omega_{m0}=0.28$,
$\Omega_{x0}=0.72$, $H_{0}$ = 72 $\pm$ 8 Kms$^{-1}$ Mpc$^{-1}$ and
a fixed prior distribution. Here we shall determine the model
parameters $A, B, C$ by minimizing the $\chi^{2}$ statistic. The
probability distribution can be written as

\begin{equation}
L= \int e^{-\frac{1}{2}{\chi}_{Stern}^{2}}P(H_{0})dH_{0}
\end{equation}

where $P(H_{0})$ is the prior distribution function for $H_{0}$.
As per our theoretical model of MCG the two parameters should
satisfy the two inequalities $A\le 1$ and $B>0$. We now plot the
graphs for different confidence levels i.e., 66\%, 90\% and 99\%
confidence levels and for three different dimensions (4D, 5D and
6D). Now our best fit analysis with Stern observational data
support the theoretical range of the parameters. When we fix the
two parameters $C=1.5$ and $\alpha=0.0001$, the 66\% (solid,
blue), 90\% (dashed, red) and 99\% (dashed, black) contours for
$(A,B)$ are plotted in figures 1, 2 and 3 for 4D $(n=2)$, 5D
$(n=3)$ and 6D $(n=4)$ respectively and we see that $A$ becomes
negative in this case. If we fix the parameter $A$ and
$\alpha=0.0001$, the 66\% (solid, blue), 90\% (dashed, red) and
99\% (dashed, black) contours for $(B,C)$ are plotted in (i)
figures 4-6 for 4D, 5D and 6D respectively with $A=1$, (ii)
figures 7-9 for 4D, 5D and 6D respectively with $A=1/3$ and (iii)
figures 10-12 for 4D, 5D and 6D respectively with $A=-1/3$. The
best fit values of $(B,C)$ and minimum values of $\chi^{2}$ for
different values of $A=1,~1/3,~-1/3$ in different dimensions are
tabulated in Table 2. For each dimension, we compare the model
parameters through the values of the parameters and by the
statistical contours. From this comparative study, one can
understand the convergence of theoretical values of the parameters
to the values of the parameters obtained from the observational
data set and how it changes from normal four dimension to higher
dimension (6D).\\\\\\

\begin{figure}
\includegraphics[scale=0.5]{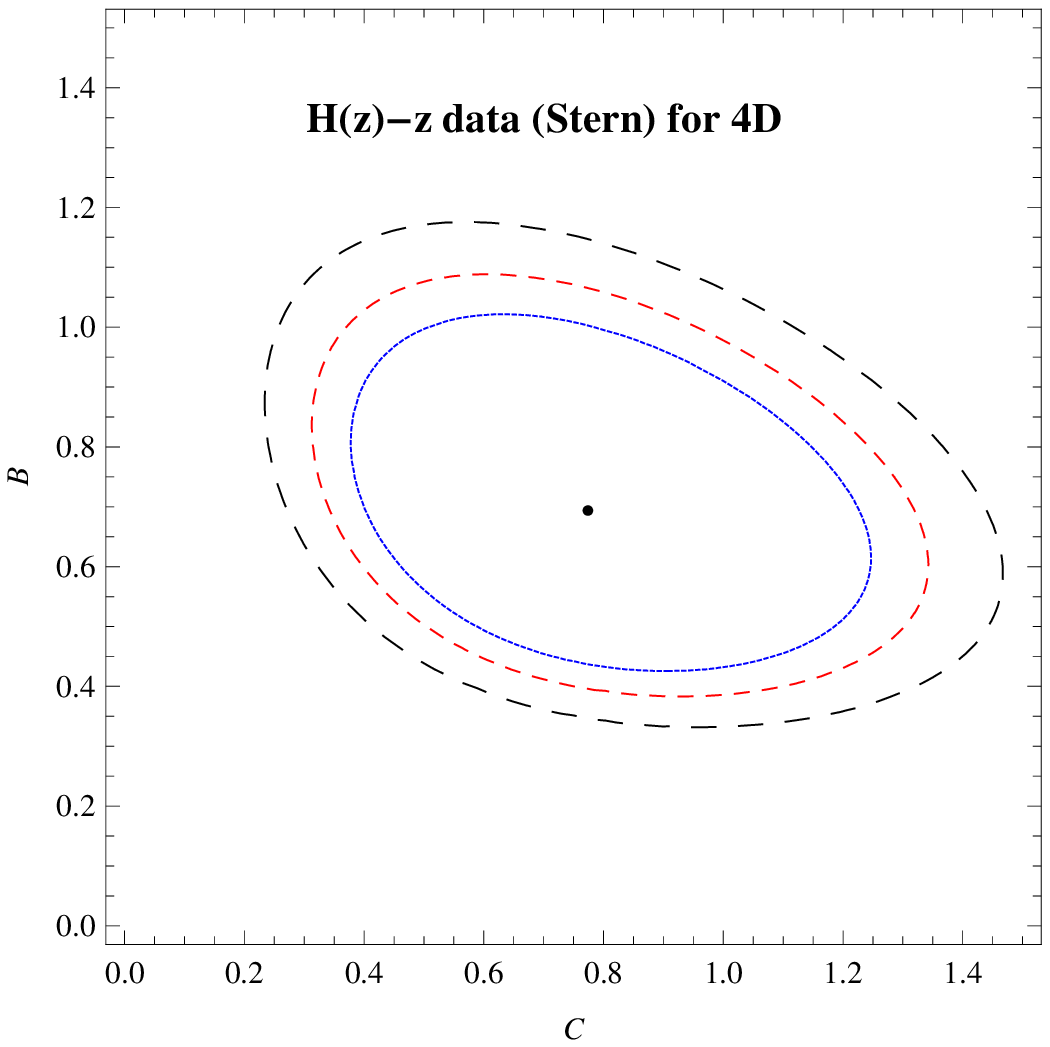}~~
\includegraphics[scale=0.5]{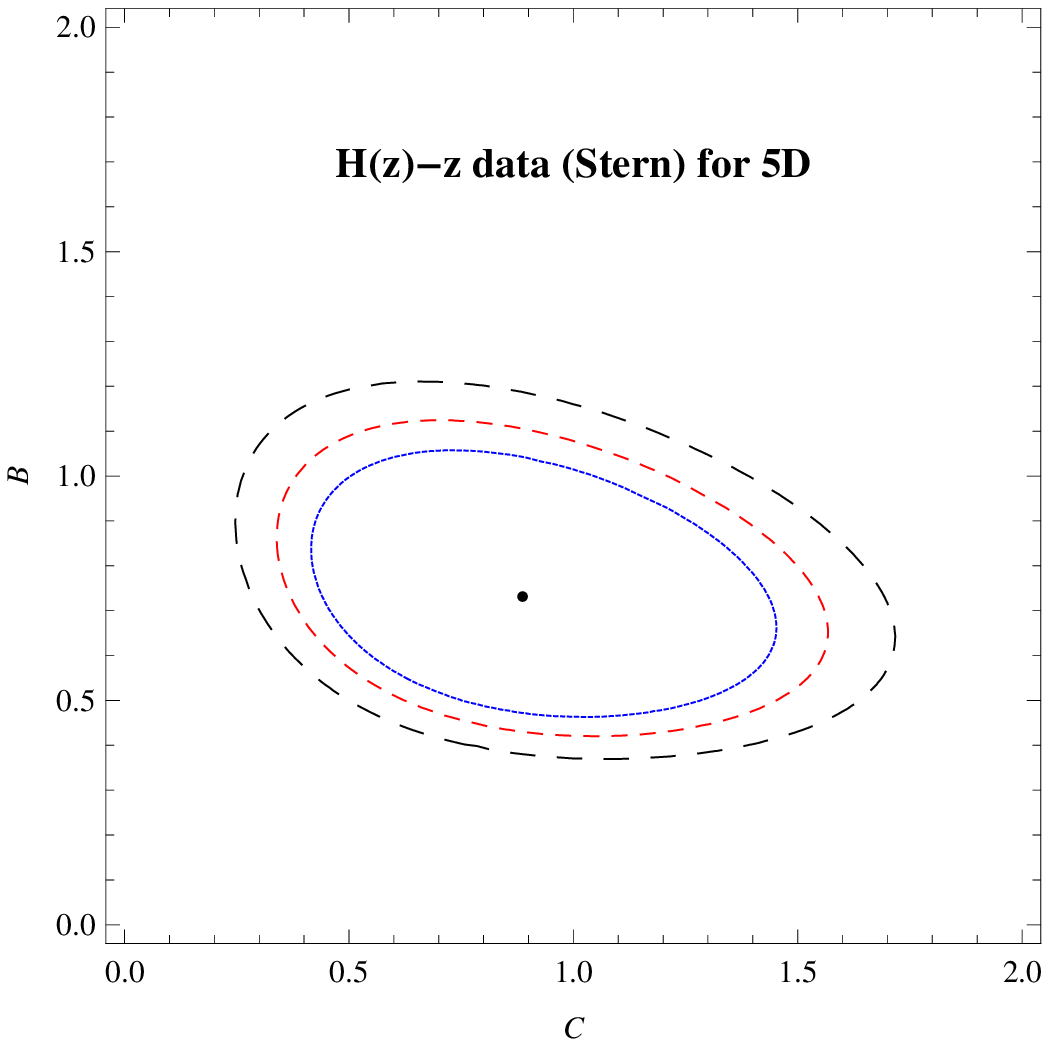}~~
\includegraphics[scale=0.5]{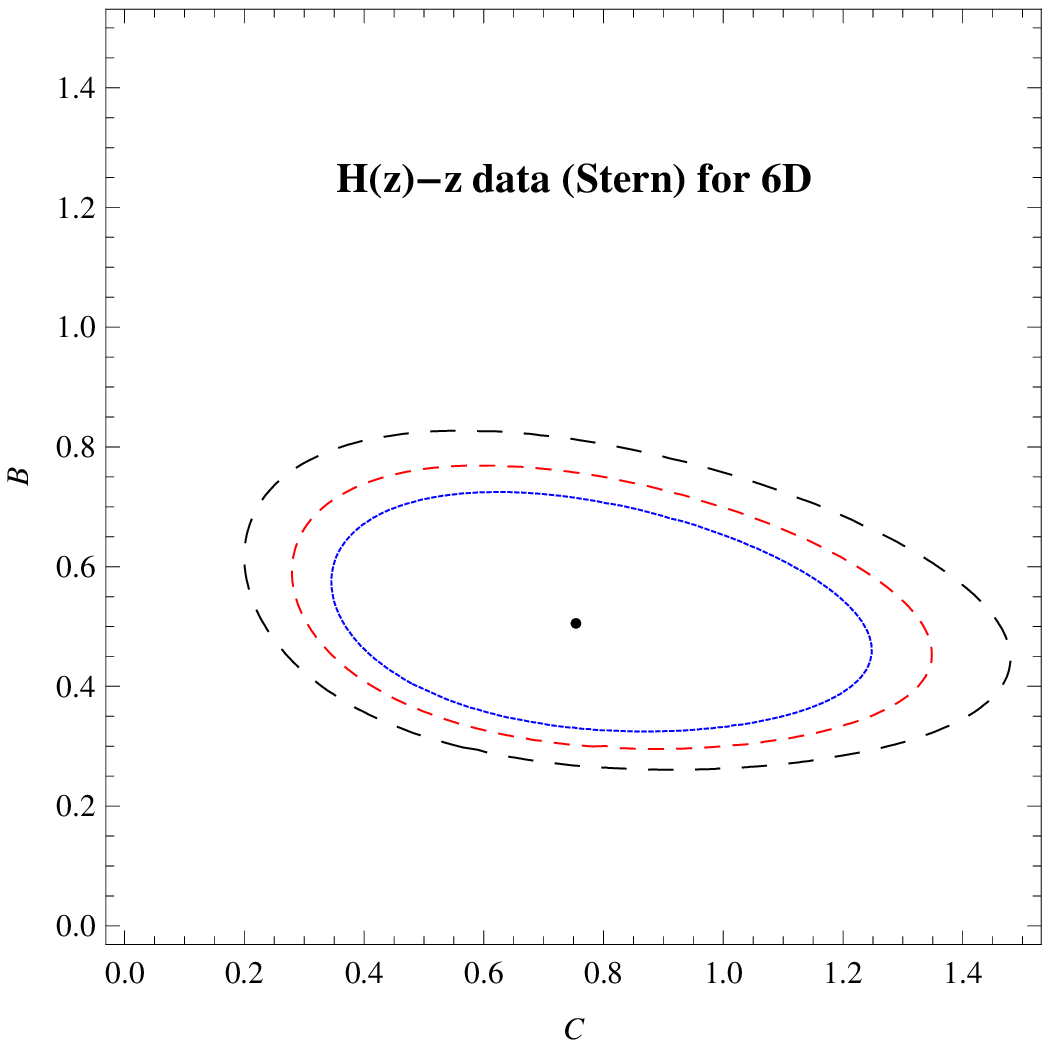}\\
\vspace{2mm}
~Fig.4~~~~~~~~~~~~~~~~~~~~~~~~~~~~~~~~~~~~~~~~~~~Fig.5~~~~~~~~~~~~~~~~~~~~~~~~~~~~~~~~~~~~~~~~~~~Fig.6\\
\vspace{4mm} $A=1$

\vspace{5mm}

\includegraphics[scale=0.5]{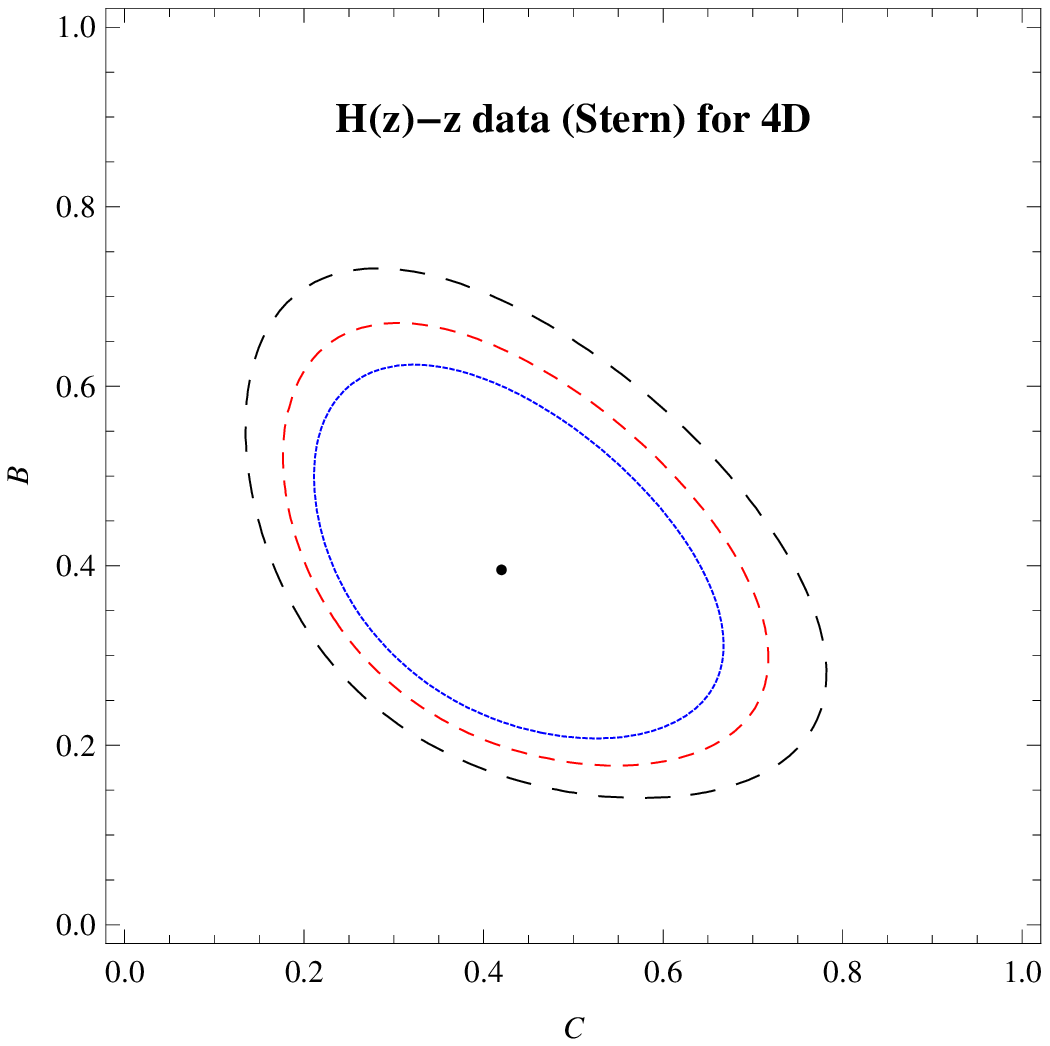}~~
\includegraphics[scale=0.5]{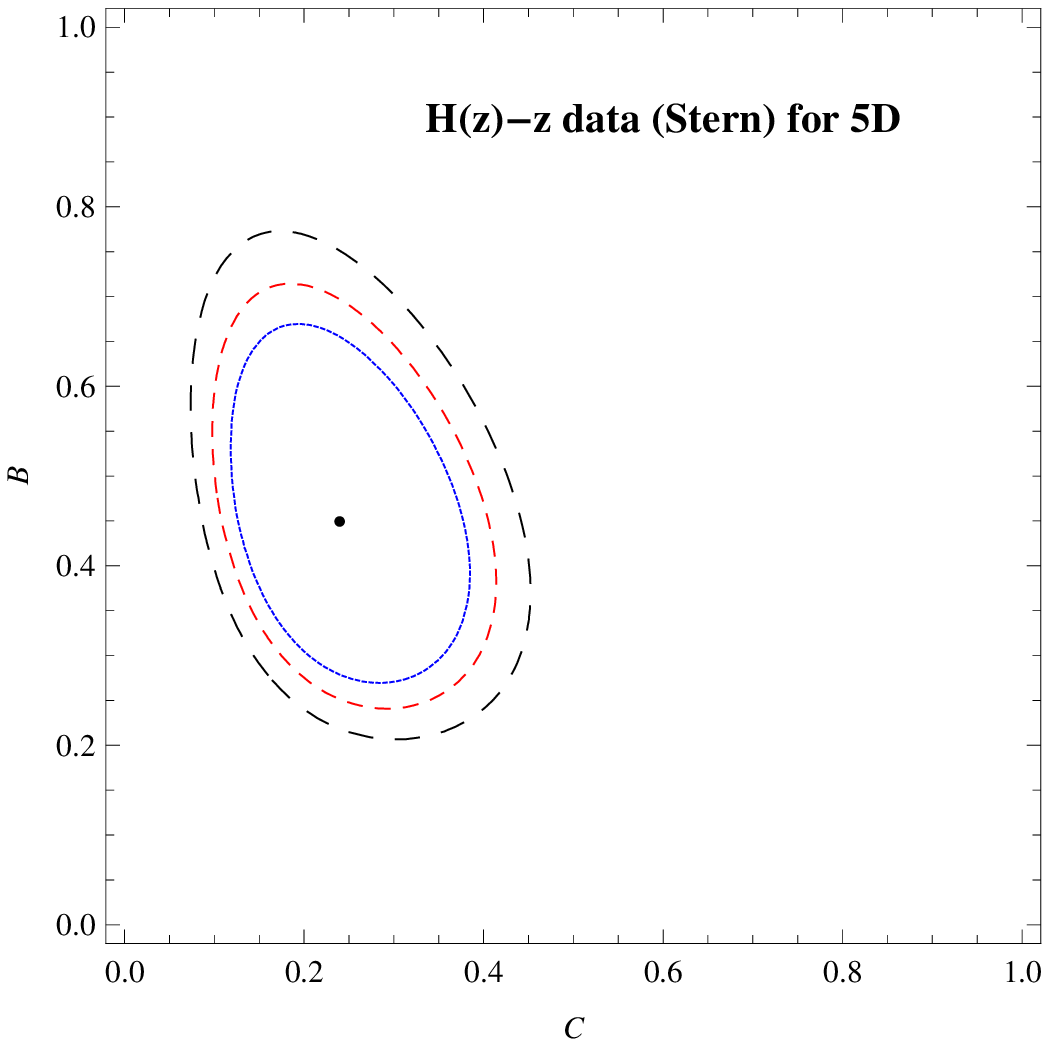}~~
\includegraphics[scale=0.5]{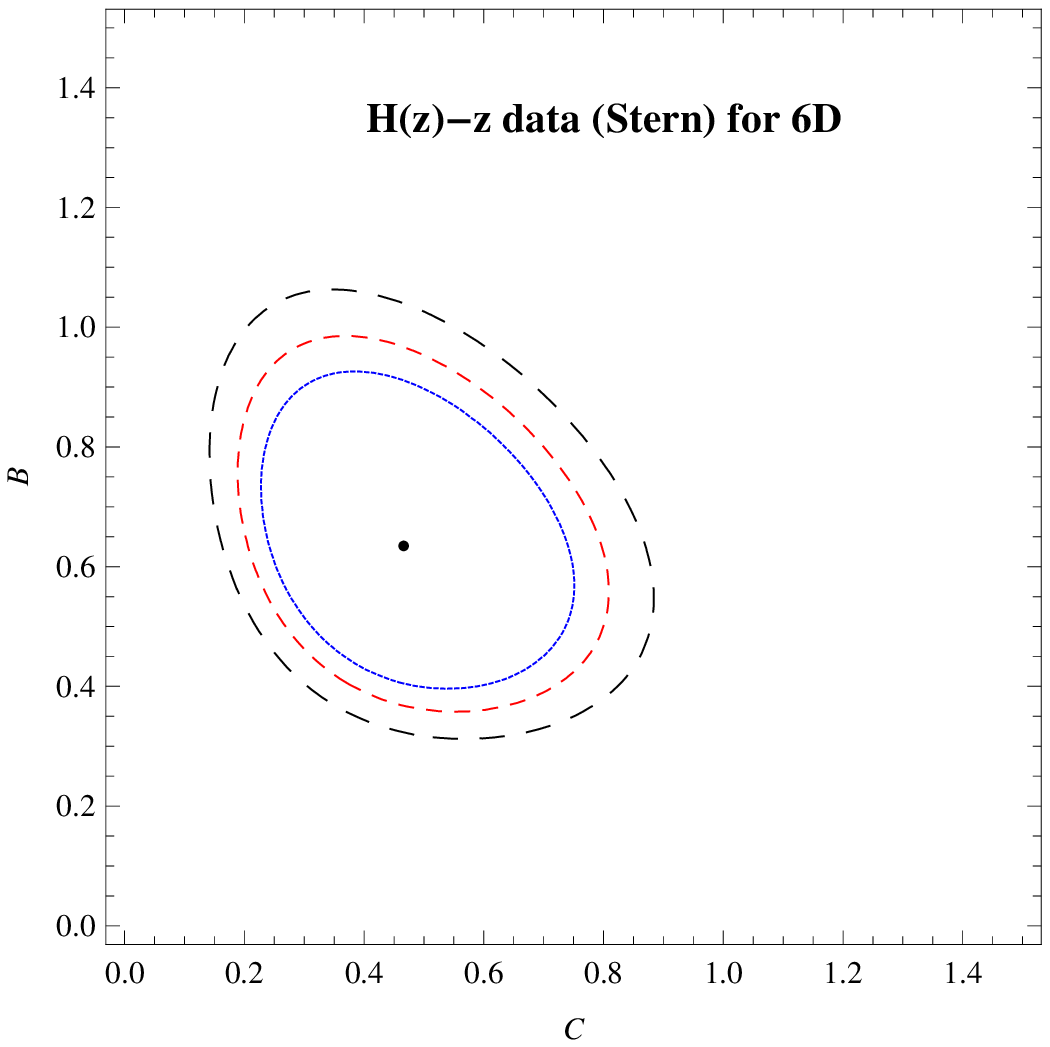}\\
\vspace{2mm}
~Fig.7~~~~~~~~~~~~~~~~~~~~~~~~~~~~~~~~~~~~~~~~~~~Fig.8~~~~~~~~~~~~~~~~~~~~~~~~~~~~~~~~~~~~~~~~~~~Fig.9\\
\vspace{4mm} $A=\frac{1}{3}$

\vspace{5mm}

\includegraphics[scale=0.5]{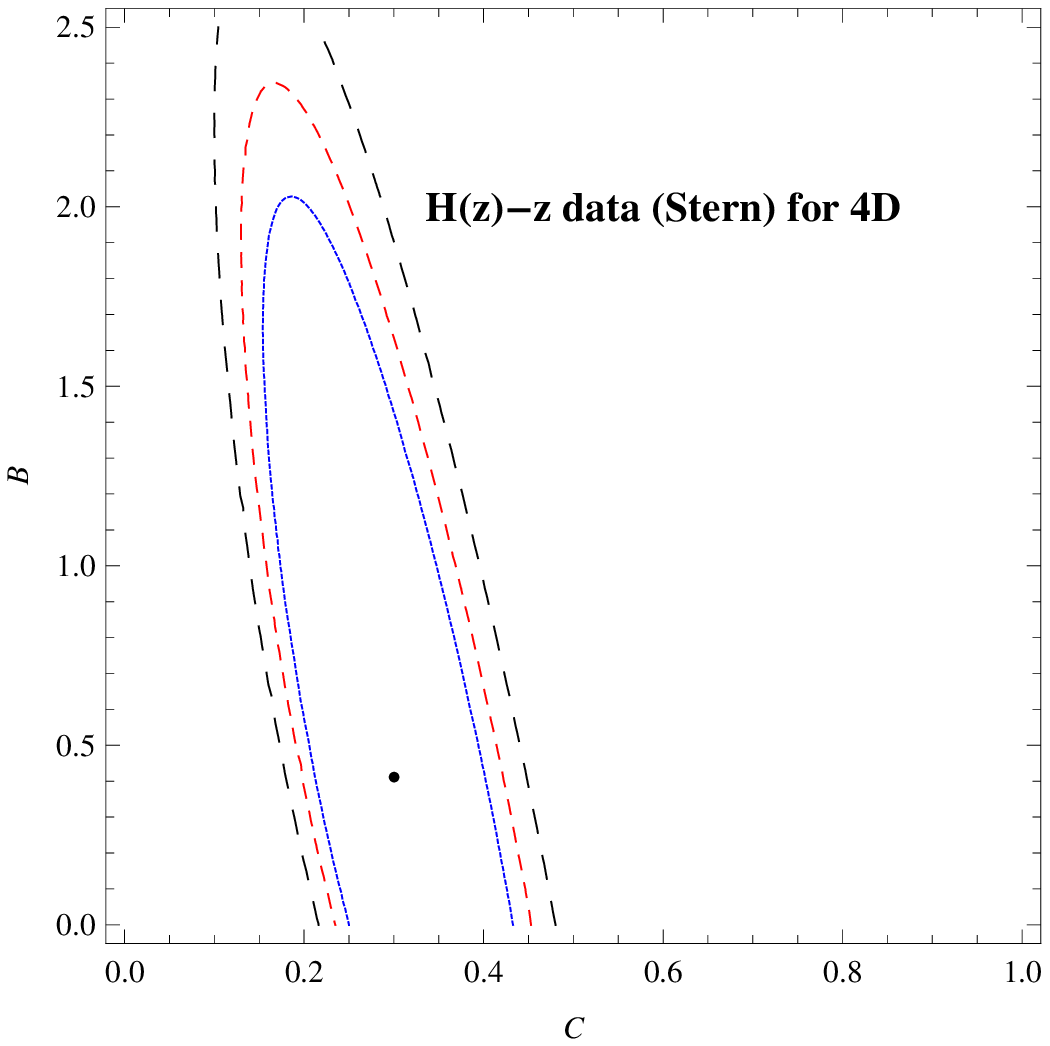}~~
\includegraphics[scale=0.5]{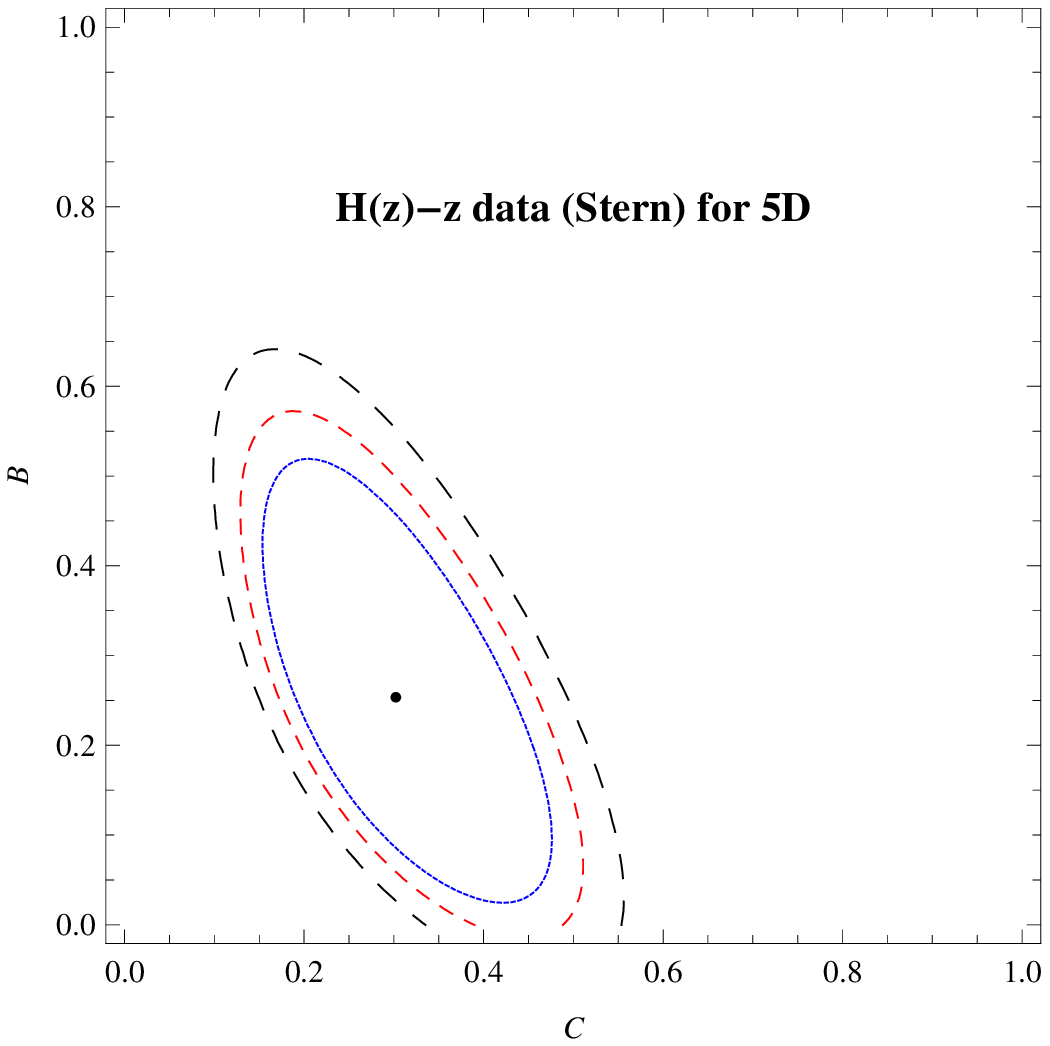}~~
\includegraphics[scale=0.5]{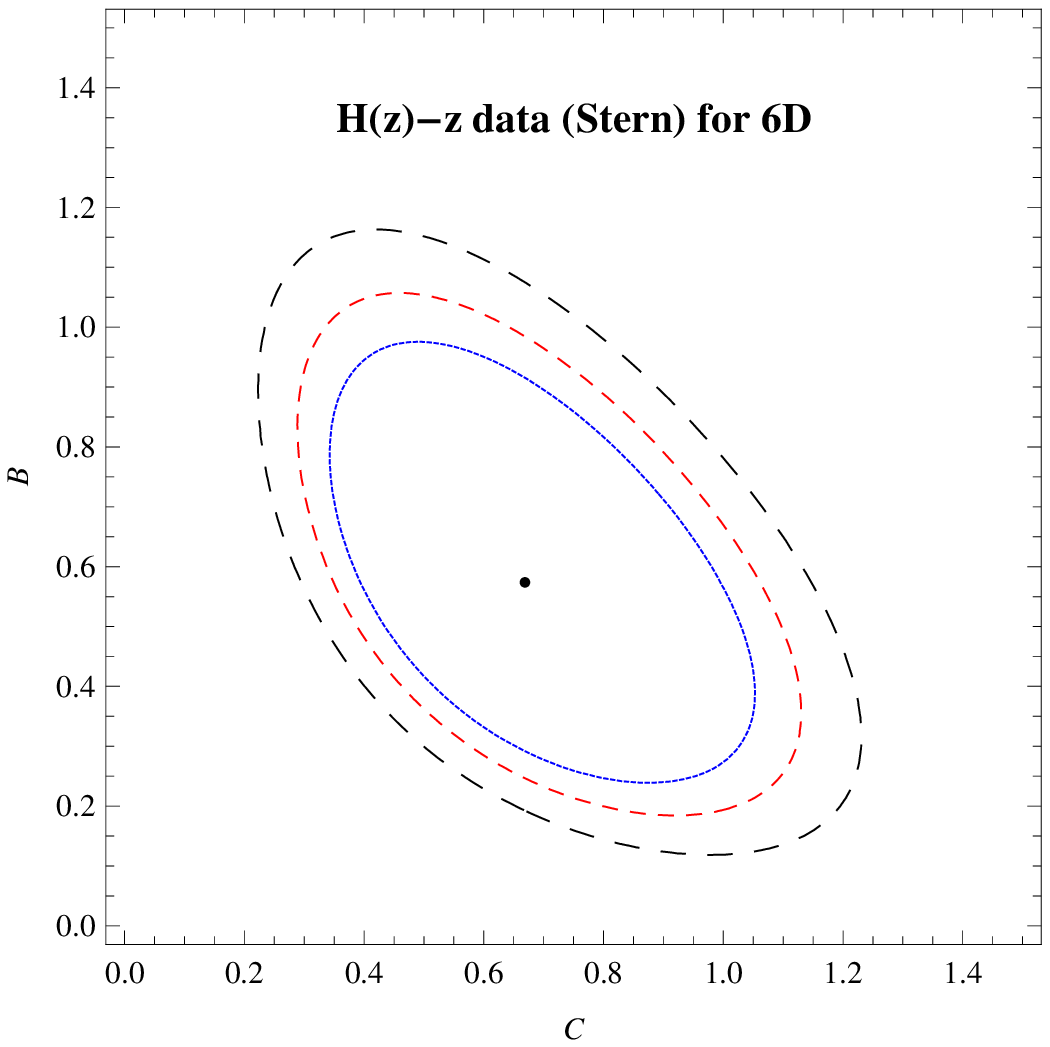}\\
\vspace{2mm}
~Fig.10~~~~~~~~~~~~~~~~~~~~~~~~~~~~~~~~~~~~~~~~~~~Fig.11~~~~~~~~~~~~~~~~~~~~~~~~~~~~~~~~~~~~~~~~~~~Fig.12\\
\vspace{4mm} $A=-\frac{1}{3}$

\vspace{4mm}

Figs. 4 - 12 show that the variation of $B$ against $C$ for
$\alpha= 0.0001$ in 4D, 5D and 6D respectively for different
confidence levels for $A=1$ (figs. 4-6), $A=1/3$ (figs. 7-9),
$A=-1/3$ (figs. 10-12). The 66\% (solid, blue), 90\% (dashed, red)
and 99\% (dashed, black) contours are plotted in these figures for
the $H(z)$-$z$ (Stern) analysis.

\vspace{4mm}
\end{figure}

\[
\begin{tabular}{|c|c|c|c|}
\hline
  ~~~~~~~~~$n$ ~~~~~~& ~~~~~~~$A$ ~~~~~~~~& ~~~$B$~~~~~&~~~~~$\chi^{2}_{min}$~~~~~~\\
  \hline
     2(4D) & $-$0.330 & 0.419 & 7.065 \\
     3(5D) & $-$0.334 & 0.511 & 7.154 \\
     4(6D) & $-$0.381 & 0.625 & 7.464 \\
  \hline
\end{tabular}
\]
{\bf Table 2:} $H(z)$-$z$ (Stern): The best fit values of $A$, $B$
and the minimum values of $\chi^{2}$ in three different dimensions
and for $\alpha=0.0001$ and $C=1.5$.

\[
\begin{tabular}{|c|c|c|c|c|}
\hline
  ~~~~~~~~$A$ ~~~~~~~~~&~~~~~~~$n$ ~~~~~~& ~~~~~~~$B$ ~~~~~~~~& ~~~$C$~~~~~&~~~~~$\chi^{2}_{min}$~~~~~~\\
  \hline
  $  $ & 2(4D) & 0.694 & 0.774 & 11.484 \\
  $~~1$ & 3(5D) & 0.731 & 0.887 & 15.114 \\
  $    $ & 4(6D) & 0.505 & 0.754 & 19.191 \\
  \hline
  $             $ & 2(4D) & 0.396 & 0.420 & 8.382 \\
  $~~\frac{1}{3}$ & 3(5D) & 0.449 & 0.240 & 10.378 \\
  $              $ & 4(6D) & 0.634 & 0.466 & 12.769 \\
  \hline
  $           $ & 2(4D) & 0.411 & 0.300 & 7.066 \\
  $-\frac{1}{3}$ & 3(5D) & 0.254 & 0.302 & 7.152 \\
  $             $ & 4(6D) & 0.574 & 0.668 & 7.632 \\
   \hline
\end{tabular}
\]
{\bf Table 3:} $H(z)$-$z$ (Stern): The best fit values of $B$, $C$
and the minimum values of $\chi^{2}$ for different values of $A$
in three different dimensions and for $\alpha=0.0001$.

\subsection{\bf{Stern + BAO Joint Data Analysis}}

The Baryon Acoustic Oscillations (BAO) in the primordial
baryon-photon fluid, leave a characteristic signal on the galaxy
correlation function, a bump at a scale $\sim$ 100 Mpc, as
observed by Eisenstein et al \cite{Eisenstein}. The peaks and
troughs seen in the angular power spectrum arise from
gravity-driven acoustic oscillations of the coupled photon-baryon
fluid in the early Universe. The interpretation of BAO
measurements are the effects of non-linear gravitational
evolution, of scale-dependent differences between the clustering
of galaxies and of dark matter and for spectroscopic surveys,
redshift distortions of the clustering, which can shift the BAO
features. Sloan Digital Sky Survey (SDSS) survey is one of the
first redshift survey (46748 luminous red galaxies spectroscopic
sample, over 3816 square-degrees of sky approximately five billion
light years in diameter) by which the BAO signal has been directly
detected at a scale $\sim$ 100 Mpc (SDSS confirmed the WMAP
results that the sound horizon in the today's universe). The SDSS
catalog provides a picture of the distribution of matter such that
one can search for a BAO signal by seeing if there is a larger
number of galaxies separated at the sound horizon. We shall
investigate the two parameters $A$ and $B$ for our model using the
BAO peak joint analysis for low redshift (with range $0 < z <
0.35$) using standard $\chi^2 $ distribution. The BAO peak
parameter may be defined by

\begin{eqnarray}
{\cal{A}}=\frac{\sqrt{\Omega_m}}{E(z_1)^{1/3}}\left(\frac{1}{z_1}~\int_{0}^{z_1}\frac{dz}{E(z)}\right)^{2/3}
\end{eqnarray}

where
\begin{eqnarray}
\Omega_m=\Omega_{m0}(1+z_1)^3 E(z_1)^{-2}
\end{eqnarray}

Here, $E(z)$ is the normalized Hubble parameter and $z_1 = 0.35$
is the typical redshift of the SDSS data sample. This quantity can
be used even for more general models which do not present a large
contribution of dark energy at early times. Now the $\chi^2$
function for the BAO measurement can be written as in the
following form

\begin{eqnarray}
\chi^2_{BAO}=\frac{({\cal{A}}-0.469)^2}{0.017^2}
\end{eqnarray}

where the value of the parameter ${\cal{A}}$ for the flat model
($k=0$) of the FRW universe is obtained by ${\cal{A}}=0.469 \pm
0.017$ using SDSS data set \cite{Eisenstein} from luminous red
galaxies survey. Now the total joint data analysis (Stern+BAO) for
the $\chi^2$ function defined by

\begin{eqnarray}
\chi^2_{Tot}=\chi^2_{Stern}+\chi^2_{BAO}
\end{eqnarray}

Now our best fit analysis with Stern$+$BAO observational data
support the theoretical range of the parameters. In figures 13, 14
and 15, we plot the graphs of $(A,B)$ for different confidence
levels 66\% (solid, blue), 90\% (dashed, red) and 99\% (dashed,
black) contours for 4D, 5D and 6D respectively and by fixing the
other parameters $C=1.5$ and $\alpha=0.0001$.

\[
\begin{tabular}{|c|c|c|c|}
\hline
  ~~~~~~~~~$n$ ~~~~~~& ~~~~~~~$A$ ~~~~~~~~& ~~~$B$~~~~~&~~~~~$\chi^{2}_{min}$~~~~~~\\
  \hline
     2(4D) & $-$0.323 & 0.440 & 767.456 \\
     3(5D) & $-$0.327 & 0.517 & 767.508 \\
     4(6D) & $-$0.373 & 0.632 & 767.787 \\
  \hline
\end{tabular}
\]
{\bf Table 4:} $H(z)$-$z$ (Stern+BAO): The best fit values of $A$,
$B$ and the minimum values of $\chi^{2}$ in three different
dimensions and for $\alpha=0.0001$ and $C=1.5$.

\begin{figure}
\includegraphics[scale=0.5]{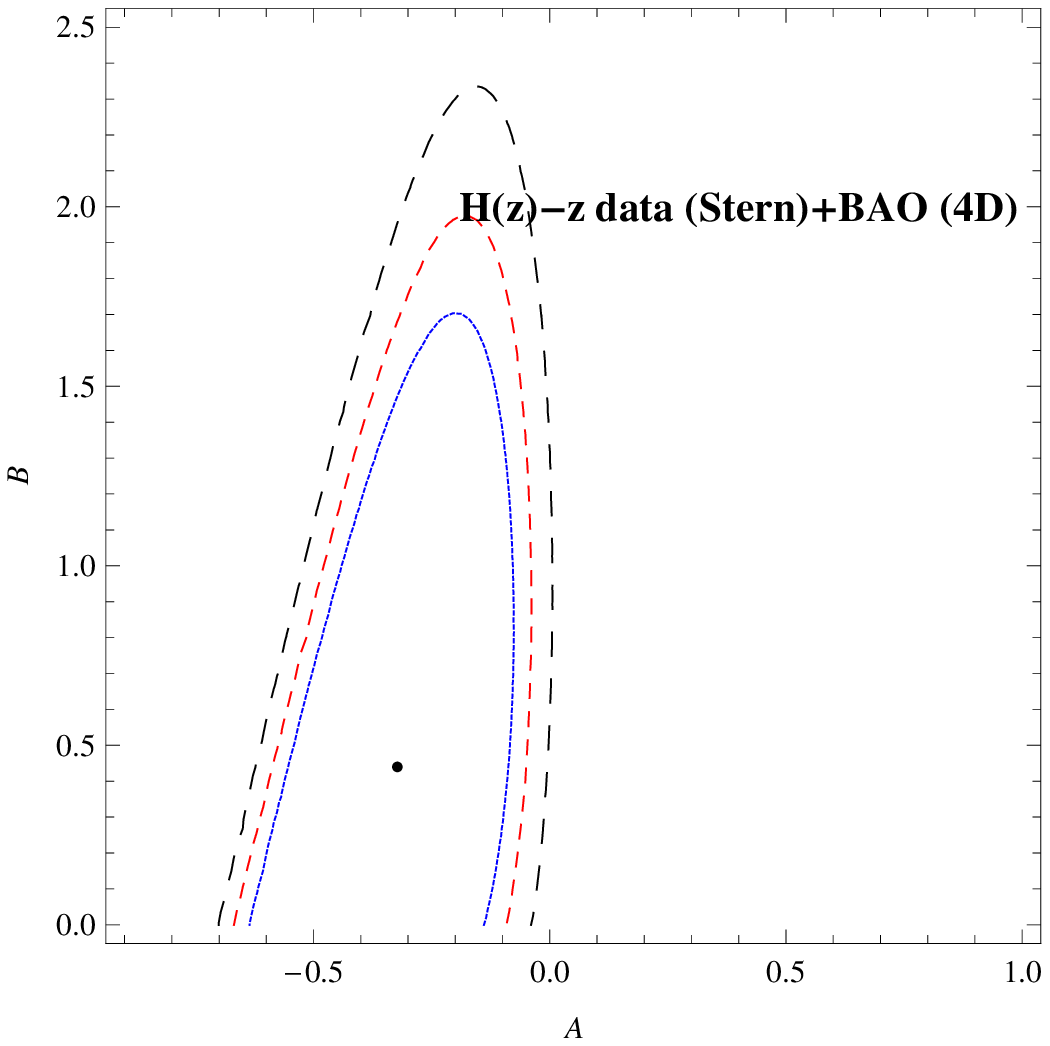}~~
\includegraphics[scale=0.5]{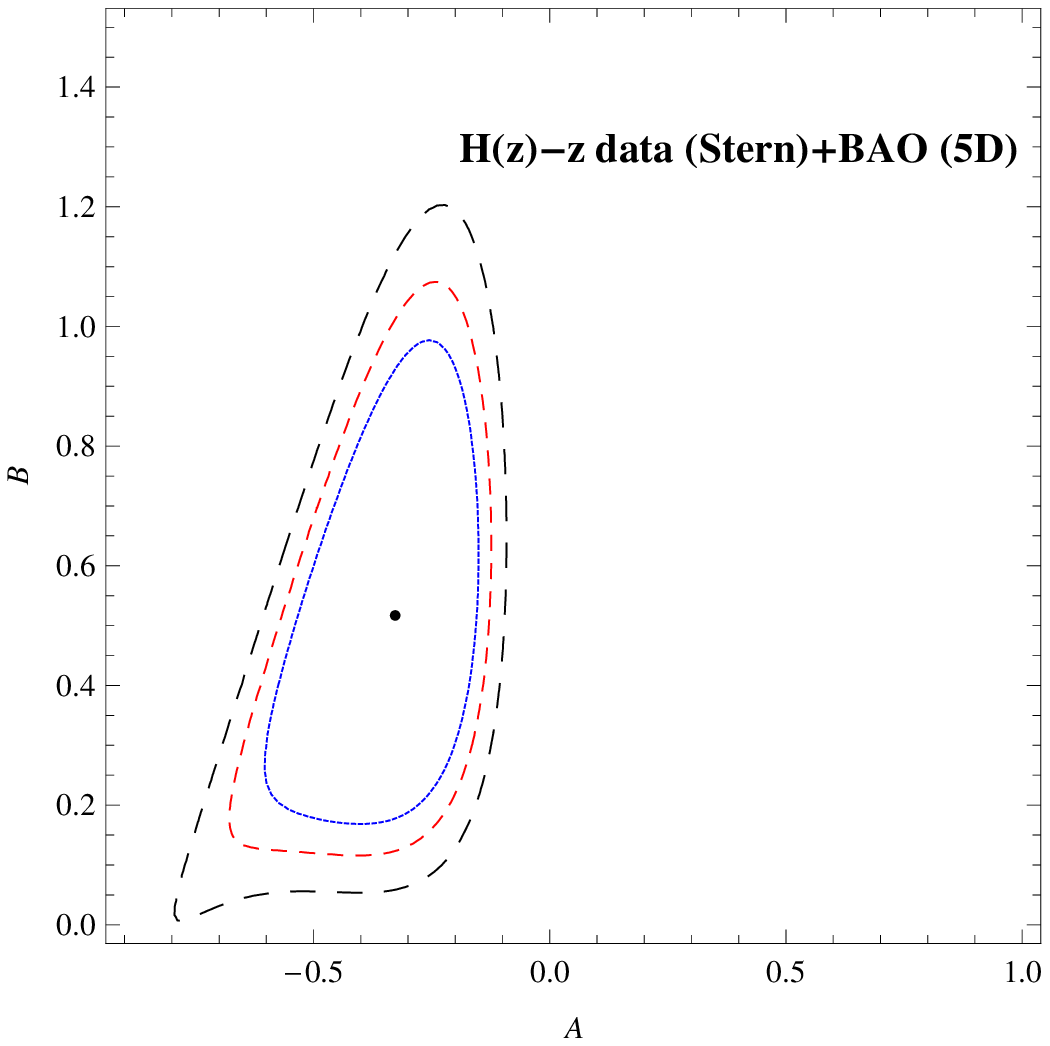}~~
\includegraphics[scale=0.5]{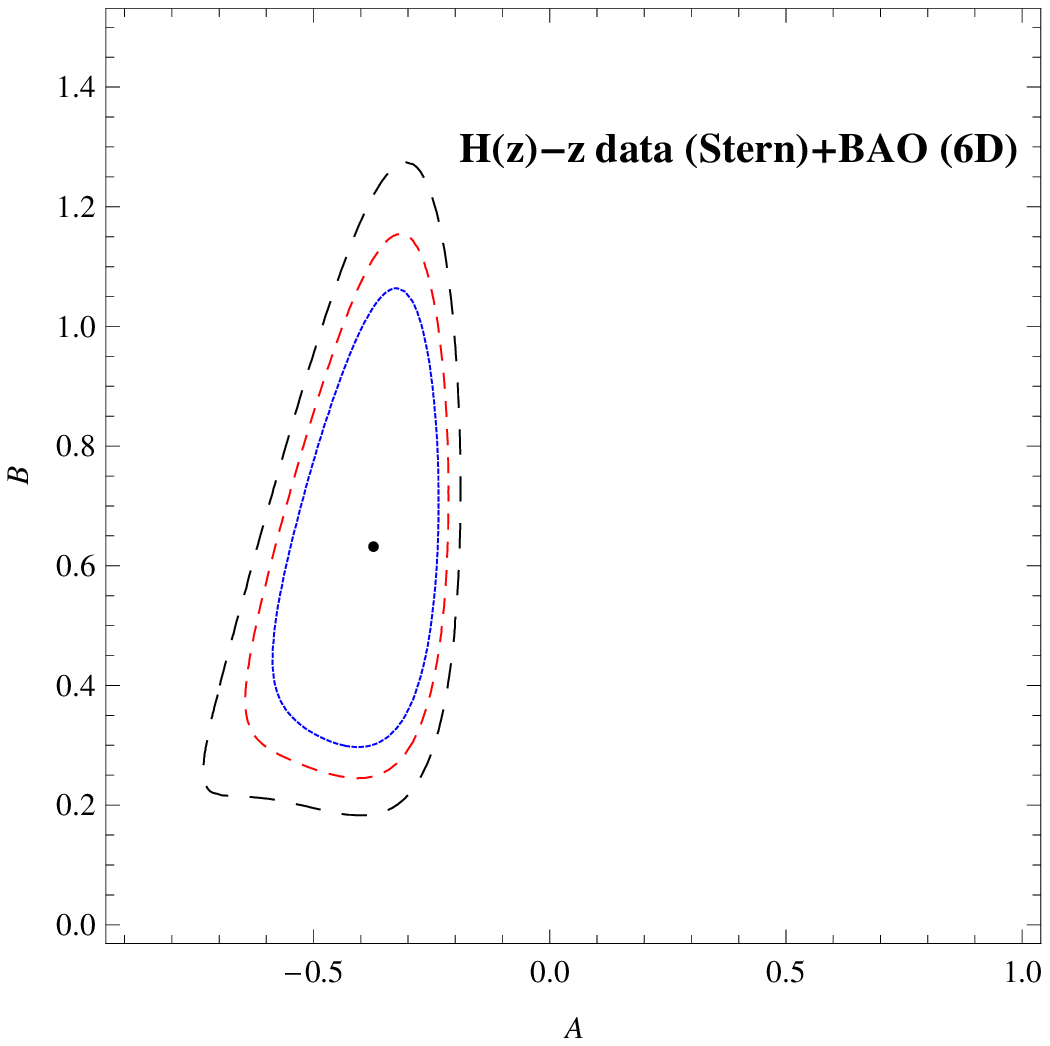}\\
\vspace{2mm}
~Fig.13 ~~~~~~~~~~~~~~~~~~~~~~~~~~~~~~~~~~~~~~~~~~~Fig.14 ~~~~~~~~~~~~~~~~~~~~~~~~~~~~~~~~~~~~~~~~~~~Fig.15 \\
\vspace{4mm}

Figs. 13 - 15 show that the variation of $B$ against $A$ for
$\alpha= 0.0001$ and $C=1.5$ in 4D, 5D and 6D respectively for
different confidence levels. The 66\% (solid, blue), 90\% (dashed,
red) and 99\% (dashed, black) contours are plotted in these
figures for the Stern+BAO analysis.

\vspace{4mm}
\end{figure}

\section{\normalsize\bf{Discussions}}

In this work, we have considered the flat FRW model of the
universe in $(n+2)$-dimensions filled with the dark matter
(perfect fluid with negligible pressure) and the modified
Chaplygin gas (MCG) type dark energy. We present the Hubble
parameter in terms of the observable parameters $\Omega_{m0}$,
$\Omega_{x0}$ and $H_{0}$ with the redshift $z$ and the other
parameters like $A$, $B$, $C$, $n$ and $\alpha$. We have chosen
the observed values of $\Omega_{m0}=0.28$, $\Omega_{x0}=0.72$ and
$H_{0}$ = 72 Kms$^{-1}$ Mpc$^{-1}$. From Stern data set (12
points), we have obtained the bounds of the arbitrary parameters
by minimizing the $\chi^{2}$ test. The best-fit values of the
parameters are obtained by 66\%, 90\% and 99\% confidence levels.
Now to find the bounds of of the parameters and to draw the
statistical confidence contour, we first fixed three parameters
$C, n, \alpha$ and then fixed the three parameters $A, n, \alpha$.
In the first case we find the bounds of $(A, B)$ and draw the
contour between them for 4D$(n=2)$, 5D$(n=3)$ and 6D$(n=4)$. In
the second case we fixed three different values of A as 1, $1/3$,
$-1/3$ to find the bounds of $(B, C)$ and draw the contour between
them. Here the parameter $n$ determines the higher dimensions and
we perform comparative study between three cases : 4D $(n=2)$, 5D
$(n=3)$ and 6D $(n=4)$ respectively. Finally due to joint analysis
with Stern+BAO observational data, we find the bounds of $(A, B)$
and draw the contour between them for 4D$(n=2)$, 5D$(n=3)$ and
6D$(n=4)$.

We have plotted the graphs for different confidence levels i.e.,
66\%, 90\% and 99\% confidence levels and for three different
dimensions (4D, 5D and 6D). Now our best fit analysis with Stern
observational data support the theoretical range of the
parameters. When we fix the two parameters $C=1.5$ and
$\alpha=0.0001$, the 66\% (solid, blue), 90\% (dashed, red) and
99\% (dashed, black) contours for $(A,B)$ are plotted in figures
1, 2 and 3 for 4D $(n=2)$, 5D $(n=3)$ and 6D $(n=4)$ respectively.
The best fit values of $(A,B)$ and minimum values of $\chi^{2}$ in
different dimensions (4D, 5D and 6D) are tabulated in Table 2 and
we see that $A$ becomes negative in this case. If we fix the
parameter $A$ and $\alpha=0.0001$, the 66\% (solid, blue), 90\%
(dashed, red) and 99\% (dashed, black) contours for $(B,C)$ are
plotted in (i) figures 4-6 for 4D, 5D and 6D respectively with
$A=1$, (ii) figures 7-9 for 4D, 5D and 6D respectively with
$A=1/3$ and (iii) figures 10-12 for 4D, 5D and 6D respectively
with $A=-1/3$. The best fit values of $(B,C)$ and minimum values
of $\chi^{2}$ for different values of $A=1,~1/3,~-1/3$ in
different dimensions are tabulated in Table 3. For each dimension,
we compare the model parameters through the values of the
parameters and by the statistical contours. From this comparative
study, one can understand the convergence of theoretical values of
the parameters to the values of the parameters obtained from the
observational data set and how it changes from normal four
dimension to higher dimension (6D). Next due to joint analysis
with Stern+BAO observational data, we have also obtained the
bounds of the parameters ($A,B$) by fixing some other parameters
$\alpha$ and $C$ for 4D, 5D and 6D. In figures 13, 14 and 15, we
have plotted the graphs of $(A,B)$ for different confidence levels
66\% (solid, blue), 90\% (dashed, red) and 99\% (dashed, black)
contours for 4D, 5D and 6D respectively and by fixing the other
parameters $C=1.5$ and $\alpha=0.0001$. The best fit values of
$(A,B)$ and minimum values of $\chi^{2}$ for Stern and BAO data in
different dimensions (4D, 5D and 6D) are tabulated in Table 4. In
summary, the conclusion of this discussion suggests that for
different dimension (4D, 5D and 6D) cosmological observation can
put upper bounds on the magnitude of the correction coming from
quantum gravity that may be closer to the theoretical expectation
than what one would expect.\\

{\bf Acknowledgement:}\\

The authors are thankful to IUCAA, Pune, India for warm
hospitality where part of the work was carried out. Also UD is
thankful to CSIR, Govt. of India for providing research project
grant (No. 03(1206)/12/EMR-II).\\


\begin{thebibliography}{99}

\bibitem[\protect\citeauthoryear{Maddox et al.}{1990}]{Maddox}Efstathiou G., Sutherland W. J. and Maddox S. J.: \textit{Nature}, \textbf{348}, 705(1990).

\bibitem[\protect\citeauthoryear{Padmanabhan et al.}{1992}]{paddy3}Padmanabhan T. and Narasimha D.:  \textit{Mon. Not. R. Astron. Soc.}, \textbf{259}, 41P(1992).

\bibitem[\protect\citeauthoryear{Efstathiou et al.}{1992}]{Bond1}Efstathiou G., Bond J. R. and White S. D. M.: \textit{Mon. Not. R. Astron. Soc.}, \textbf{258}, 1P(1992).

\bibitem[\protect\citeauthoryear{Bagla et al.}{1996}]{Paddy4}Bagla J. S., Padmanabhan T. and Narlikar J. V.: \textit{Comments Astrophys.}, \textbf{18}, 275(1996).

\bibitem[\protect\citeauthoryear{Perlmutter et al.}{1998}]{Perlmutter}Perlmutter, S. J. et al.: \textit{Nature}, \textbf{391}, 51(1998).

\bibitem[\protect\citeauthoryear{Perlmutter et al.}{1999}]{Perlmutter1}Perlmutter, S. J. et al.: \textit{Astrophys. J.}, \textbf{517}, 565(1999).

\bibitem[\protect\citeauthoryear{Riess et al.}{1998}]{Riess}Riess, A. G. et al.: \textit{Astron. J.}, \textbf{116}, 1009(1998).

\bibitem[\protect\citeauthoryear{Riess et al.}{2004}]{Riess1}Riess, A. G. et al.: \textit{Astron. J.}, \textbf{607}, 665(2004).

\bibitem[\protect\citeauthoryear{Bachall et al.}{1999}]{Bachall}Bachall, N. A. et al.: \textit{Science}, \textbf{284}, 1481(1999).

\bibitem[\protect\citeauthoryear{Tedmark et al.}{2004}]{Tedmark}Tedmark, M. et al.: \textit{Phys. Rev. D}, \textbf{69}, 103501(2004).

\bibitem[\protect\citeauthoryear{Miller et al.}{1999}]{Miller}Miller, D. et al.: \textit{Astrophys. J.}, \textbf{524}, L1(1999).

\bibitem[\protect\citeauthoryear{Bennet et al.}{2000}]{Bennet}Bennet, C. et al.: \textit{Phys. Rev. Lett.}, \textbf{85}, 2236(2000).

\bibitem[\protect\citeauthoryear{Briddle et al.}{2003}]{Briddle}Briddle, S. et al.: \textit{Science}, \textbf{299}, 1532(2003).

\bibitem[\protect\citeauthoryear{Spergel et al.}{2003}]{Spergel}Spergel, D. N. et al.: \textit{Astrophys. J. Suppl.}, \textbf{148}, 175(2003).

\bibitem[\protect\citeauthoryear{Padmanabhan}{2003}]{Paddy}Padmanabhan, T.: \textit{Phys. Rept.} \textbf{380}, 235(2003).

\bibitem[\protect\citeauthoryear{Sahni et al.}{2000}]{Sahni}Sahni, V. and Starobinsky, A. A.: \textit{Int. J. Mod. Phys. D}, \textbf{9}, 373(2000).

\bibitem[\protect\citeauthoryear{Peebles et al.}{1988}]{Peebles}Peebles, P. J. E. and Ratra, B.: \textit{Astrophys. J. Lett.}, \textbf{325}, L17(1988).

\bibitem[\protect\citeauthoryear{Riess et al.}{2007}]{Riess2}Riess, A. G. et al.: \textit{Astrophys. J.}, \textbf{659}, 98(2007).

\bibitem[\protect\citeauthoryear{Spergel et al.}{2007}]{Spergel1}Spergel., D. N. [WMAP collaboration]: \textit{Astrophys. J. Suppl.}, \textbf{170}, 337(2007).

\bibitem[\protect\citeauthoryear{Padmanabhan et al.}{2003}]{Paddy2}Padmanabhan, T. and Choudhury, T. R.: \textit{Mon. Not. R. Astron. Soc.}, \textbf{344}, 823(2003).

\bibitem[\protect\citeauthoryear{Tonry et al.}{2003}]{Tonry}Tonry, J. L. et al.: \textit{ApJ}, \textbf{594}, 1(2003).

\bibitem[\protect\citeauthoryear{Barris et al.}{2004}]{Barris}Barris, B. J. et al., \textit{ApJ}, \textbf{602}, 571(2004).

\bibitem[\protect\citeauthoryear{Choudhury et al.}{2007}]{Paddy1}Choudhury, T. R. and Padmanabhan, T.: \textit{Astron. Astrophys.}, \textbf{429}, 807(2007).

\bibitem[\protect\citeauthoryear{Kamenshchik et al.}{2001}]{Kamenshchik}Kamenshchik, A. et al.: \textit{Phys. Lett. B}, \textbf{511}, 265(2001).

\bibitem[\protect\citeauthoryear{Gorini et al.}{2003}]{Gorini}Gorini, V.,  Kamenshchik, A. and Moschella, U.: \textit{Phys. Rev. D}, \textbf{67}, 063509(2003).

\bibitem[\protect\citeauthoryear{Bento et al.}{2002}]{Bento}Bento, M. C.,  Bertolami, O. and Sen, A. A.: \textit{Phys. Rev. D}, \textbf{66}, 043507(2002).

\bibitem[\protect\citeauthoryear{Debnath et al.}{2004}]{Debnath}Debnath, U., Banerjee, A. and Chakraborty, S.: \textit{Class. Quantum Grav.}, \textbf{21}, 5609(2004).

\bibitem[\protect\citeauthoryear{Lu et al.}{2008}]{Lu}Lu, J. et al, \textit{Phys. Lett. B}, \textbf{662}, 87(2008).

\bibitem[\protect\citeauthoryear{Jun et al.}{2005}]{Jun}Dao-Jun, L. and Xin-Zhou, L.: \textit{Chin. Phys. Lett.}, \textbf{22}, 1600(2005).

\bibitem[\protect\citeauthoryear{Poppenhaeger et al.}{2004}]{Poppenhaeger}Poppenhaeger, K., Hossenfelder, S., Hofmann, S. and Bleicher, M.: \textit{Phys. Lett. B}, \textbf{582}, 1(2004).

\bibitem[\protect\citeauthoryear{Arkani et al.}{1998}]{Arkani}Arkani-Hamed, N., Dimopoulos, S. and Dvali, G.: \textit{Phys. Lett. B}, \textbf{429}, 263(1998).

\bibitem[\protect\citeauthoryear{Panigrahi et al.}{2008}]{Panigrahi1}Panigrahi, D. and Chatterjee, S.: \textit{Gen. Rel. Grav.}, \textbf{40}, 833(2008).

\bibitem[\protect\citeauthoryear{Chatterjee}{2009}]{Chatterjee}Chatterjee, S.: \textit{arXiv: 0911.2621} [gr-qc].

\bibitem[\protect\citeauthoryear{Mukhopadhyay}{2010}]{Utpal}U. Mukhopadhyay, P. P. Ghosh and S. Ray, arXiv:1001.0475 [gr-qc].

\bibitem[\protect\citeauthoryear{Stern et al.}{2010}]{Stern}Stern, D. et al.: \textit{JCAP}, \textbf{1002}, 008(2010).

\bibitem[\protect\citeauthoryear{Wu et al.}{2007}]{Wu}Wu, P. and Yu, H.: \textit{Phys. Lett. B}, \textbf{644}, 16(2007).

\bibitem[\protect\citeauthoryear{Thakur et al.}{2009}]{Paul}Thakur, P., Ghose, S. and Paul, B. C.: \textit{Mon. Not. R. Astron. Soc.}, \textbf{397}, 1935(2009).

\bibitem[\protect\citeauthoryear{Paul et al.}{2011}]{Paul1}Paul, B. C., Ghose, S. and Thakur, P.: \textit{arXiv: 1101.1360v1} [astro-ph.CO].

\bibitem[\protect\citeauthoryear{Paul et al.}{2010}]{Paul2} Paul, B. C., Thakur, P. and Ghose, S.: \textit{arXiv: 1004.4256v1}[astro-ph.CO].

\bibitem[\protect\citeauthoryear{Ghose et al.}{2011}]{Paul3} Ghose, S., Thakur, P. and Paul, B. C.: \textit{arXiv: 1105.3303v1}[astro-ph.CO].

\bibitem[\protect\citeauthoryear{Chakraborty et al.}{2012}]{Chak}Chakraborty, S., Debnath, U. and Ranjit, C.: \textit{Eur. Phys. J. C}, \textbf{72}, 2101(2012).

\bibitem[\protect\citeauthoryear{Eisenstein et al.}{2005}]{Eisenstein}Eisenstein, D. J. et al.: \textit{Astrophys. J.}, \textbf{633}, 560(2005).




\end{thebibliography}
\end{document}